\documentclass[12pt]{article}
\usepackage{latexsym,amsmath,amssymb,amsbsy,graphicx}
\usepackage[utf8]{inputenc}
\usepackage[T2A]{fontenc}
\usepackage[space]{cite} 

\makeatletter\def\@biblabel#1{\hfill#1.}\makeatother
\allowdisplaybreaks
\begin {document}

\noindent\begin{minipage}{\textwidth}
\begin{center}

{\Large{On the Effectiveness of Observations in the Mid-Infrared Wavelength Range on the 2.5-Meter Telescope of the Caucasus Mountain Observatory of Moscow State University with Commercial IR Cameras}}\\[9pt]

\textbf{S.\,G. Zheltoukhov$^{1,2}$, A.\,M. Tatarnikov$^{1,2a}$}\\[6pt]

\parbox{.96\textwidth}{\centering\small
\textit {$^1$Sternberg Astronomical Institute,
Moscow State University, Moscow 119234, Russia.}\\
\textit{$^2$Department of Experimental Astronomy,
Faculty of Physics, M.V.Lomonosov Moscow State University, Moscow 119991, Russia.}\\
\ E-mail: $^a$andrew@sai.msu.ru}\\[1cc] 

\parbox{.96\textwidth}{\centering\small Received July 11, 2022; revised October 10, 2022; accepted October 12, 2022} 
\end{center}

{\parindent5mm The main factors that influence the success of observations in the infrared range (central wavelengths of the photometric bands at 3.75 and 4.8~$\mu$m) on the multipurpose optical telescope are considered. Estimates of the sky background brightness are obtained for the Caucasus Mountain Observatory (CMO) of Moscow State University: $1.3\cdot10^6$~photons/(s pixel) in the 3.75~$\mu$m band and
$3.4\cdot10^6$~photons/(s pixel) in the 4.8~$\mu$m; and the instrumental background for the 2.5-m CMO telescope at $0^\circ$С: $3.2\cdot10^6$~photons/(s pixel) in the 3.75~$\mu$m band and $4.3\cdot10^6$~photons/(s pixel) in the 4.8~$\mu$m band. It is shown that at this background signal level with the currently available commercial cameras in the $3-5$~$\mu$m spectral range, the telescope-camera coupling capabilities for observing faint objects will still be limited by the thermal background. For different observational conditions, estimates of the limiting magnitudes of objects available for observations in the 3.75 and 4.8~$\mu$m ranges are obtained. For average observation conditions (instrument temperature of $0^\circ$С and stellar image size of $1''$), the limit is $\sim10.6^m$ and $\sim8.4^m$, respectively.
\vspace{2pt}\par}

\vspace{1pt}\par

\vspace{1pt}\par
\end{minipage}

\section*{Introduction}
\mbox{}\vspace{-\baselineskip}

Infrared (IR) radiation is electromagnetic radiation over a wide range of wavelengths~--- from 0.8 to 300-500~$\mu$m. The infrared range borders the visual range on the short wavelength side and the microwave range on the long wavelength side. For convenience, this range is divided into several sub-ranges. One of the most commonly used splits is as follows: near-infrared range with boundaries of 0.8~- 2.5~$\mu$m, mid-infrared range with boundaries of 2.5~- 25~$\mu$m, far-infrared range with boundaries of 25~- 500~$\mu$m. In the near-infrared range, it is possible to carry out astronomical observations with ordinary non-adapted reflector telescopes, in the mid-infrared range to reduce the instrumental background it is necessary to apply various design solutions (a cooled Lyot stops, no telescope baffles, small secondary mirror size, radiation flux modulation, etc.; the longer the wavelength, the more adaptation of the telescope is required), in the far IR, observations are possible only from a spacecraft or high-altitude airplane.

At present, there are not many specialized IR telescopes in operation in the world, on which observations in the mid-IR range are possible. In Russia, there is only one such telescope~--- AZT-33IR \cite{kamus2002}, on which observations are currently carried out in the optical wavelength range of 0.3-1.1~$\mu$m. In the IR in Russia, regular observations are carried out on only two non-adapted telescopes~--- in the 1-5~$\mu$m range at the 1.25-m telescope ZTE of the Crimean Astronomical Station of SAI (with a single-element photometer \cite{Nadjip1986}, \cite{Shenavrin2011}) and in the 1-2.5~$\mu$m range at the 2.5-m telescope of the Caucasus Mountain Observatory of SAI (CMO, \cite{Shatsky2020}) with a camera-spectrograph ASTRONIRCAM \cite{Nadjip2017}.

In recent years, due to the progress of IR detection technology, the cost of IR detectors and IR cameras have tended to decrease. At the same time, the characteristics are improving and the spectral range accessible to commercially available serial products is increasing. The purpose of the present work is to estimate the background level and determine the limiting magnitudes of the current unadapted for IR observations 2.5-m telescope of CMO when operated with a commercial IR detector at wavelengths from 3 to 5~$\mu$m and to present the design of a IR camera prototype.

\section{Problem statement}
\mbox{}\vspace{-\baselineskip}

CMO is located 20~km south of Kislovodsk (Russia) at an altitude of 2100~m. Astroclimate monitoring performed by Kornilov et al. in 2007~-2013 \cite{Kornilov2014} showed that for the CMO, the median value of the FWHM of stars (Full Width at Half Magnitude~--- width of the star's image profile at the level of half of the maximum value) in the visible wavelength range is 0.96~arcsec, the median value of precipitated water column for clear nights is $PWV=7.8$~mm, and the greatest number of clear nights falls in the fall-winter period. The relatively low atmospheric water content (especially in late fall and winter) and the high altitude of the observatory above sea level make observations in the infrared possible. The main instrument of the observatory~--- 2.5-m telescope, which has been operating in test mode since 2014 and will be officially commissioned at the end of 2021.

The 2.5-m telescope of CMO is a universal multitasking telescope that allows us to perform a wide range of studies. It is built according to the optical scheme of the classical Ritchey-Chretien reflector (Table ~\ref{table:telescope_parameters}) with an additional tertiary Nasmyth mirror M3. Movement and rotation of the M3 mirror help to realize five different positions of the focal plane: the Cassegrain focus (the M3 mirror is out of the beam) and the four Nasmyth focuses (the M3 mirror is introduced into the beam and tilted at an angle of $45^\circ$ to the optical axis of the main mirror), realized by rotating the M3 mirror around the optical axis of the telescope's main mirror. It is planned that the mid-infrared camera will be located at one of the Nasmyth focuses. From a camera design point of view, all these focuses are equivalent.

At present, there are detectors and cameras of the mid-IR of several vendors on the market costing 50k-70k EUR. This is an order of magnitude (or more) lower than the cost of scientific detectors with export restrictions. They use InSb or HgCdTe matrix as a sensor, and cooling to temperatures of $<100$K is carried out by a Stirling machine. However, not all existing even commercial cameras are available for ordering in Russia. Therefore, for further consideration, we will take average parameters of available commercial ones. As will be shown below, an important property of the detector is its sensitivity in the wavelength range required for astronomical observations. We will assume that the used detector operates in the entire wavelength range (i.e., from 3 to 5~$\mu$m) with a sharp fall in sensitivity at the long-wavelength boundary.

The parameters that determine the possibility of observing bright objects or working at high background level are the minimum accumulation time and the charge capacity of a pixel. Characteristic values of these parameters are $t_{min}<1$~ms and $P_e\sim8\cdot10^6$~e-. The parameters determining the angular field of view and angular resolution of the camera for astronomical observations on a particular telescope are the size of the detector sensitive area and the pixel size $a$. Let us set the first value to the often encountered format of 640x512 pixels, and for the second~--- $a=15$~µm.

We set the basic parameters of the entire path (Earth's atmosphere, telescope, and detector) through which the radiation of observed astronomical objects passes. Based on these parameters, we investigate the limitations imposed by them on the accuracy of photometric observations and evaluate the limits of combining a telescope not adapted to IR observations with a commercial mid-IR camera.

\begin{table}[htbp]
\caption{Main parameters of the 2.5-m telescope of CMO}
\begin{center}
\begin{tabular}{| l | l |}
\hline
Parameter & Value \\
\hline
Optical scheme & Ritchey-Chretien \\
Main mirror diameter $D$ & 2.5~m \\
Relative aperture $A$ & 1/8 \\
Linear central obscuration $d$ & 0.408 \\
Mirror reflectivity & 0.95 \\
Diameter of the star image, where 80\% of the energy is collected& $0.3''$ \\
Image Scale & $10''/$мм \\
Maximum pointing speed & 3$^\circ$/с \\
Maximum pointing acceleration  & 1$^\circ$/с$^2$\\
\hline
\end{tabular}
\label{table:telescope_parameters}
\end{center}
\end{table}

\section{Estimation of sky background brightness}
\mbox{}\vspace{-\baselineskip}

\begin{figure}[h]
  \center{
  \includegraphics[width=0.6\linewidth]{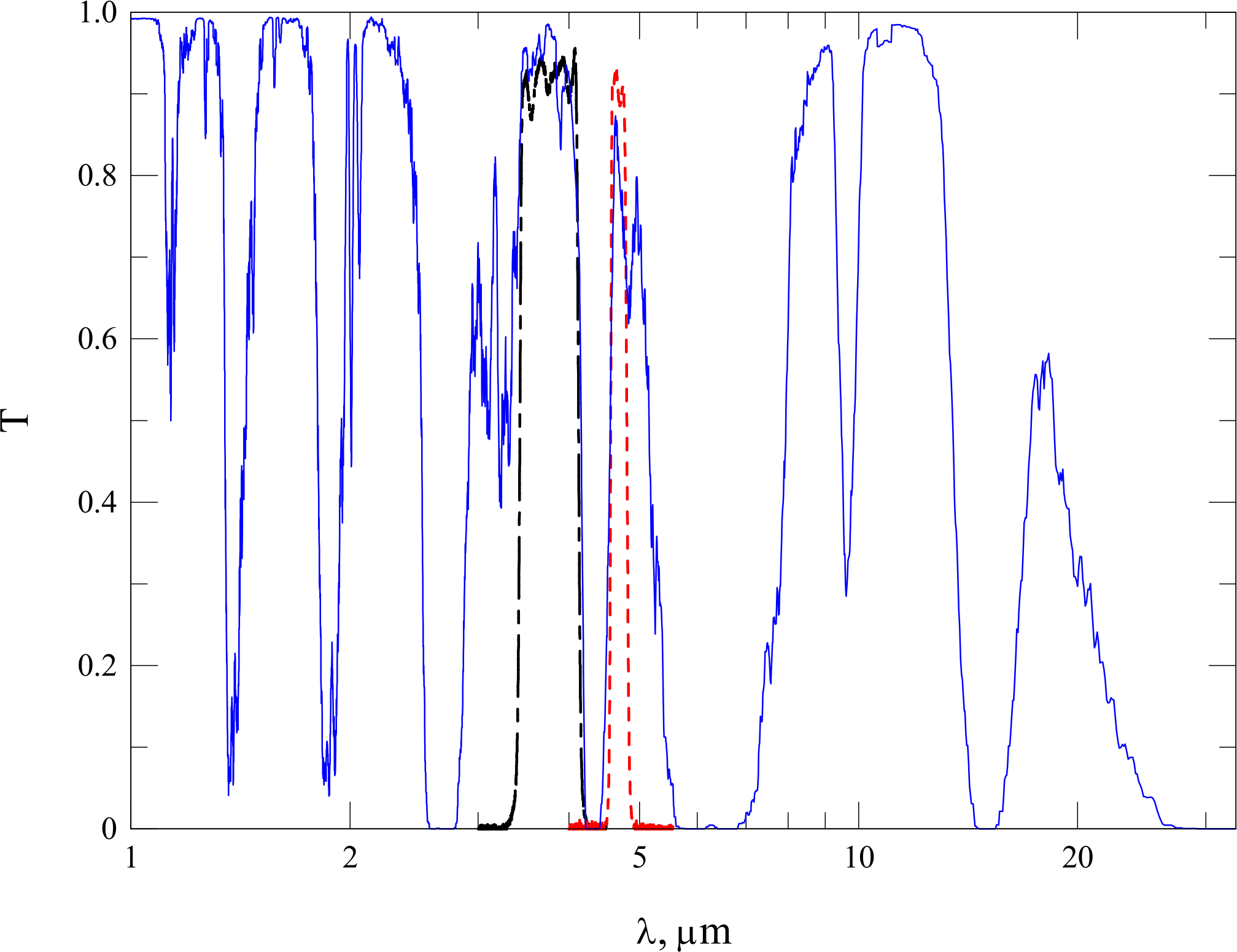}}
  \caption{Dependence of the transmission of the Earth's atmosphere on wavelength (for altitude 2000~m above the Earth's surface and PWV=1~mm) and the transmission curves of the $L'$ (dashed-dotted line) and $M'$ (dashed line) filters}
  \label{fig:atm_smooth}
\end{figure}

One of the main factors interfering with ground-based observations in the IR range is the influence of the Earth's atmosphere. The smoothed characteristic curve of atmospheric transmission in the IR range is shown in Fig.~\ref{fig:atm_smooth}. Up to a wavelength of 20~$\mu$m, we can see the alternation of absorption and transmission bands (called windows of atmospheric transparency). The main absorbing agents are water vapor, carbon dioxide, and ozone. Their concentration in the atmosphere is not constant, which causes the shape of absorption bands near their boundaries to change.

CMO is located at an altitude of 2100 m above sea level. For IR observations in the conditions of the mountain observatory, a broadband photometric system MKO-NIR (see \cite{MKO1} and\cite{MKO2}) with bands $J, H, K, L', M'$ was developed in the early 2000s. The reaction curves of the system are chosen in such a way that the strong absorption bands of water vapor and carbon dioxide, as well as the variable boundaries of these absorption bands, are avoided as far as possible. ASTRONIRCAM IR camera operates in the near-IR range (bands $J, H, K$ of the MKO-NIR system) on the 2.5-m telescope of CMO. Therefore, the MKO-NIR system was also chosen to realize the mid-IR photometric bands with the new camera. The parameters of the $L'$ and $M'$ filters are given in Table ~\ref{table:filters_parameters}, and their response curves are shown in Fig.~\ref{fig:atm_smooth}. It can be seen that for the $M'$ band, it was not possible to achieve complete independence of the system response curve from the atmospheric transmission.

\begin{table}[h]
\caption{Main parameters of the $L'$ and $M'$ filters}
\begin{center}
\begin{tabular}{| l | l | l |}
\hline
Parameter & $L'$ & $M'$ \\
\hline
Central wavelength, $\mu$m & 3.75 & 4.70 \\
Bandwidth  & 0.7 & 0.21 \\
(at 50\% of the peak), $\mu$m &   &   \\
Peak transmission & 0.9 & 0.9 \\
Transmission outside the workig range & $< 0.005$ & $< 0.005$ \\
\hline
\end{tabular}
\label{table:filters_parameters}
\end{center}
\end{table}

\begin{figure}[tbh!]
  \centerline{
  \includegraphics[width=0.6\linewidth]{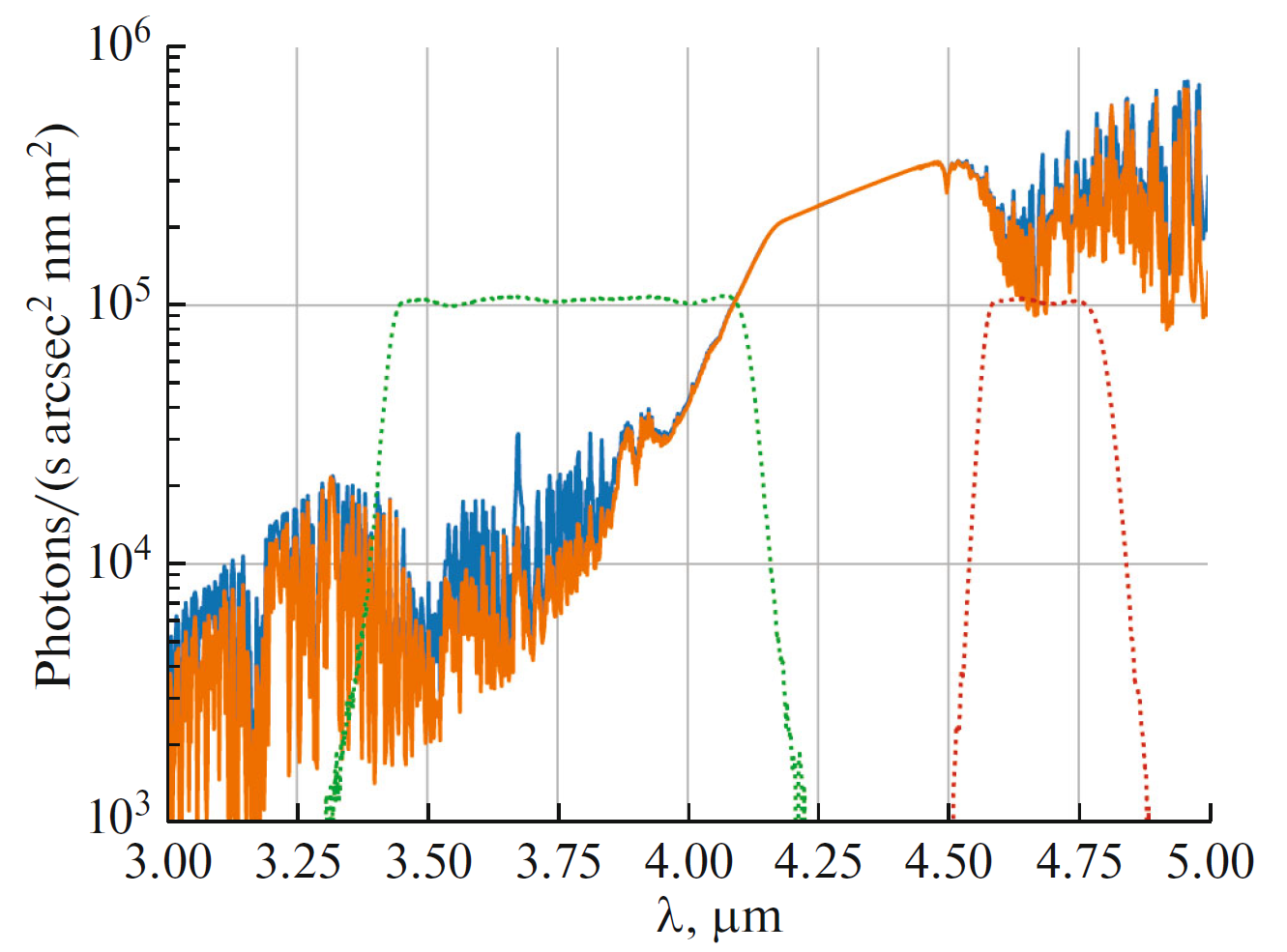}}
  \caption{Sky background spectrum from Gemini website (https://www.gemini.edu/observing/telescopes-and-sites/sites\#Near-IR-long) for airmass 1.5. Blue curve~--- sky background for PWV=10~mm. Orange curve~--- for PWV=2.3~mm. The dotted lines show the transmission curves of the $L'$ and $M'$ filters}
  \label{fig:sky_bg}
\end{figure}

In the infrared range, the Earth's atmosphere is a powerful source of background radiation. The value of the radiation flux depends strongly on the wavelength~--- if in the $K$ band (2.2~$\mu$m) the surface brightness of the sky is $13. 1^m$ per square arcsec (measured from observations at CMO), then in the $L'$ band it is $4^m$, and in the $M'$ band it is $1^m$ \footnote[1]{https://www.gemini.edu/observing/telescopes-and-sites/sites\#Near-IR-long memorandum \cite{Lord1992} was used to calculate the data on this page. }. The corresponding values in terms of energy are $F_K=3\cdot10^{-5}$ W/m$^2$/sr, $F_{L'}=4\cdot10^{-2}$ W/m$^2$/sr, $F_{M'}=8\cdot10^{-2}$ W/m$^2$/sr. The brightness in the $L'$ and $M'$ bands depend on PWV~---as the amount of precipitated water increases from 2~mm to 10~mm, the brightness of the sky background in these bands changes by about 20\% \footnotemark[1]. In Fig.~\ref{fig:sky_bg} shows a plot of the dependence of the number of photons emitted by 1 square arcsec of sky on wavelength. By convolving the above data with the transmission curves of the $L'$ and $M'$ filters, we can obtain the photon flux in them determined by the background: for the $L'$ band at the 2.5-meter telescope of CMO (see Table \ref{table:telescope_parameters}) the flux is $1.3\cdot10^6$ photons/s, for the $M'$ flux is $3.4\cdot10^6$ photons/s.

The accuracy of observations is affected not only by the high brightness of the sky background but also by its rapid fluctuations that are typical for the mid-infrared range. To account for fluctuations, it is necessary to simultaneously (quasi-simultaneously) observe both the background and the object, which is possible due to the high frame rate of the IR detectors used.

\section{Estimation of instrumental background}
\mbox{}\vspace{-\baselineskip}

In the mid-IR and far-IR, the thermal radiation of the telescope and some parts of the equipment strongly affects the obtained images. Starting from about 2~$\mu$m wavelength, the power of thermal radiation of a telescope not adapted to IR observations becomes comparable to the power of the background radiation of the sky. Below are calculations of the photon flux falling on the detector from the telescope and structural elements of the camera for different temperatures.

\subsection{Telescope Radiation}
\mbox{}\vspace{-\baselineskip}

\begin{figure}[tbh!]
  \centerline{
  \includegraphics[width=0.6\linewidth]{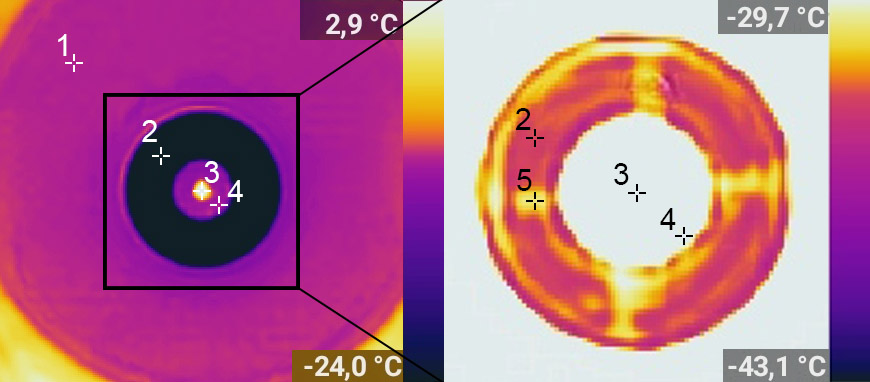}}
  \caption{Thermal map of the inner parts of the telescope obtained on December 24, 2021 at an air temperature outside the dome of $-15^\circ$C. An enlarged fragment of the inner part of the image is shown on the right; the color scale is different in the images. Point 1 ($-7.2^\circ$C)~--- the M3 mirror baffle, point 2 ($-32.9^\circ$C)~--- the sky visible after reflection from the three mirrors of telescope, point 3 ($+7.2^\circ$C)~--- the reflection of a warm observer with a thermal imager in the telescope's secondary mirror, point 4 ($-8. 1^\circ$C)~--- reflection of the inner part of the M3 mirror baffle in the secondary mirror, point 5 ($-30.5^\circ$C)~--- temperature measured near the image of the metal spider on which the secondary mirror is attached}
  \label{fig:termo_map}
\end{figure}

To determine the contribution of the telescope radiation to the instrumental background, we measured the thermal radiation from the telescope elements using a thermal imager. The operating wavelength range of the thermal imager is 7-14~µm, which allows us to see the intrinsic thermal radiation, which is most intense in this range. The thermal imager was placed in the same location where the IR camera will be installed in the future. As a result, we obtained a heat map for the internal parts of the telescope that will eventually fall within the camera's field of view. The corresponding image can be seen in Fig.~\ref{fig:termo_map}.

To a first approximation, the emissivity coefficient of the non-mirror elements of the telescope can be considered close to 1 both at the operating wavelengths of the thermal imager and in the $L'$ and $M'$ bands. In this case, the temperature data obtained with the thermal imager and the angular dimensions of the telescope elements visible from the detector are unambiguously converted to the signal seen by the detector. Let us estimate the number of photons $N_{ph}$ coming from an element with temperature $T$ and angular area $\Omega$ per detector pixel with linear size $a$ in some wavelength range:

\begin{equation}
N_{ph} = \Omega a^2 \int^{\lambda_2}_{\lambda_1} \frac{\lambda}{h c} \frac{2 h c^2}{\lambda^5} \frac{1}{\exp\left(\frac{h c}{\lambda k T}\right) - 1}  d \lambda
 \label{eq:plank_telescope}
\end{equation}

The total solid angle that a detector pixel must see during observations is determined by the relative telescope aperture $A$, and for a 2.5-meter telescope of CMO it is 41 square degrees. Of these, 35 square degrees~--- the area of relatively cold sky visible in the telescope (or more precisely~--- its reflections from the telescope mirrors) and at least 6 square degrees will be occupied by the warm elements of the telescope. Since the telescope is not optimized for IR observations, the detector's field of view includes the radiation from the primary mirror rim and the secondary mirror baffle, as well as the spider on which the secondary mirror is attached (see Fig.~\ref{fig:termo_map}).

From the data presented in Fig.~\ref{fig:termo_map}, we can get an estimate of the emissivity of the mirrors of a 2.5-meter telescope. To do this, we need to determine the ''temperature'' of the sky observed without the telescope and the sky seen through the telescope. After disabling the internal calibrations of the thermal imager, which take into account the influence of the distance to the object and the radiation reflected from it, we can consider that the flux received by the thermal imager is directly proportional to the integral of Planck's formula within the sensitivity of the thermal imager (from 7 to 14 microns) with the body temperature equal to that given by the thermal imager. Let us denote this integral by $B_{f}(T)$. We can now make an equation for the total emissivity of the mirrors, $\varepsilon$. Measurements showed that the ''temperature'' of the sky was $T_s=-39^\circ$C and the ''temperature'' of the sky through the telescope~--- $T_t=-32.9^\circ$. At the same time, sensors placed directly on the surface of the mirror showed that its temperature $T_m=-5^\circ$C. Thus we obtain the equation \ref{eq:mirrors}, from which we obtain the value $\varepsilon=0.14$, which is quite plausible for 3 consecutive mirrors.

The final results of the flux calculations from the telescope can be seen in Table ~\ref{table:telescope}.

\begin{equation}
B_{f}(T_t) = (1 - \varepsilon)B_{f}(T_s) + \varepsilon B_{f}(T_m)
 \label{eq:mirrors}
\end{equation}

\begin{table}[htbp]
\caption{Background created by the telescope}
\begin{center}
\begin{tabular}{| l | l | l | }
\hline
Temperature of the telescope, $^\circ$C & Signal in the $L'$ band, ph/s/pix & Signal in the $M'$ band, ph/s/pix  \\
\hline
-20 & 470 000 & 1 000 000\\
-10 & 820 000 & 1 600 000\\
0 & 1 400 000 & 2 500 000\\
10 & 2 200 000 & 3 700 000\\
20 & 3 500 000 & 5 300 000\\
\hline
\end{tabular}
\label{table:telescope}
\end{center}
\end{table}

\begin{figure}[tbh!]
  \centerline{
  \includegraphics[width=0.6\linewidth]{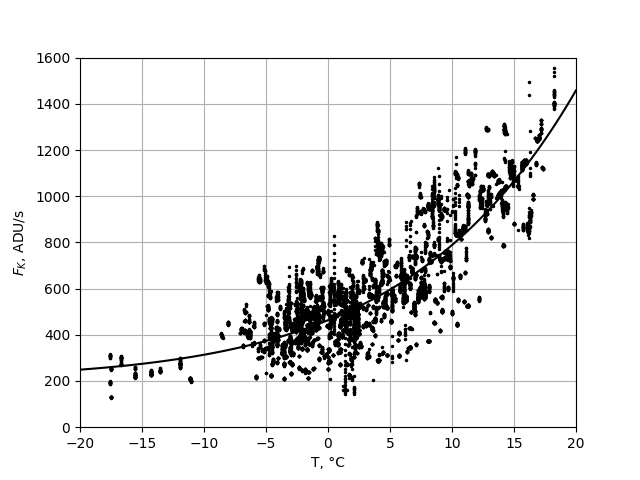}}
  \caption{Dependence of the total background value in the $K$ band on the temperature of the telescope mirror. The line shows the model dependence used in further calculations}
  \label{fig:anc_background}
\end{figure}

To verify these estimates, we measured the telescope's contribution to the background radiation using an independent method. For this purpose, we used observations made with the ASTRONIRCAM near-infrared camera \cite{Nadjip2017} has already been installed on the telescope. At wavelength >2~$\mu$m, the contribution of the telescope thermal emission becomes noticeable against the background of other sources. Therefore, we derived the dependence of the background signal recorded in the $K$ band (2.2~$\mu$m) on the telescope temperature (Fig.~\ref{fig:anc_background}). We assume that the brightness of the sky background depends weakly on the temperature of the near-surface air layer (due to the high atmospheric transparency outside the absorption bands). This makes it possible to separate the contribution to the total background flux from the sky and from the telescope. By approximating the obtained points with a model dependence, we obtain that the signal from the sky in each element of the ASTRONIRCAM detector is approximately 210 counts per second, and from the telescope at 0$^\circ$C~--- 250 counts per second. The filters and structural elements of the ASTRONIRCAM camera are cooled to cryogenic temperatures, so here we neglect their radiation. Next, with the knowledge of the photoelectron-to-counts conversion factor (gain=2.2) and the quantum efficiency of the ASTRONIRCAM camera detector \cite{Nadjip2017}, we can determine the number of $N_K$ photons falling in the $K$ band per unit time per unit area of the detector. Considering the radiation to be blackbody, we can calculate the flux coming from the telescope and in the filters $L'$ and $M'$ of interest. To do this, multiply the obtained value $N_K$ by the ratio of integrals from the Planck function for the corresponding spectral ranges and recalculate the flux by the pixel size $a$ of the detector under study (see formula \ref{eq:plank_from_k}).

\begin{equation}
N_{LM} = N_K \cdot a^2 \frac{\int^{\lambda_2}_{\lambda_1} \frac{\lambda}{h c} \frac{2 h c^2}{\lambda^5} \frac{1}{\exp\left(\frac{h c}{\lambda k T}\right) - 1}  d \lambda}{\int^{\lambda_{K2}}_{\lambda_{K1}} \frac{\lambda}{h c} \frac{2 h c^2}{\lambda^5} \frac{1}{\exp\left(\frac{h c}{\lambda k T}\right) - 1}  d \lambda}
 \label{eq:plank_from_k}
\end{equation}

After calculating these values, we obtain data that coincide well with the results presented in the table~\ref{table:telescope}, which were calculated in a different way. However, it should be taken into account that even small changes in the initial data can change the calculated background level by a factor of almost 2, i.e., the obtained results can be considered as a characteristic estimate. Nevertheless, the coincidence of the results obtained with a high probability means a good plausibility of the models used and a correct estimate of the radiative angular area of the telescope in the first calculation method.

\subsection{Radiation from photometric filters}
\mbox{}\vspace{-\baselineskip}

In addition to the radiation from various parts of the telescope, an important source of the background signal is the structural elements of the infrared camera itself. First of all, it is the entrance filter as an element occupying the largest part of the field of view. A formula similar to the \ref{eq:plank_telescope} was used to calculate the intrinsic emission of the filters:

\begin{equation}
E_{ph} =\Omega_T a^2 \int^{\lambda_2}_{\lambda_1} \varepsilon(\lambda)\frac{\lambda}{h c} \frac{2 h c^2}{\lambda^5} \frac{1}{\exp\left(\frac{h c}{\lambda k T}\right) - 1}  d \lambda
 \label{eq:plank_filter}
\end{equation}

\begin{figure}[tbh!]
  \centerline{
  \includegraphics[width=0.6\linewidth]{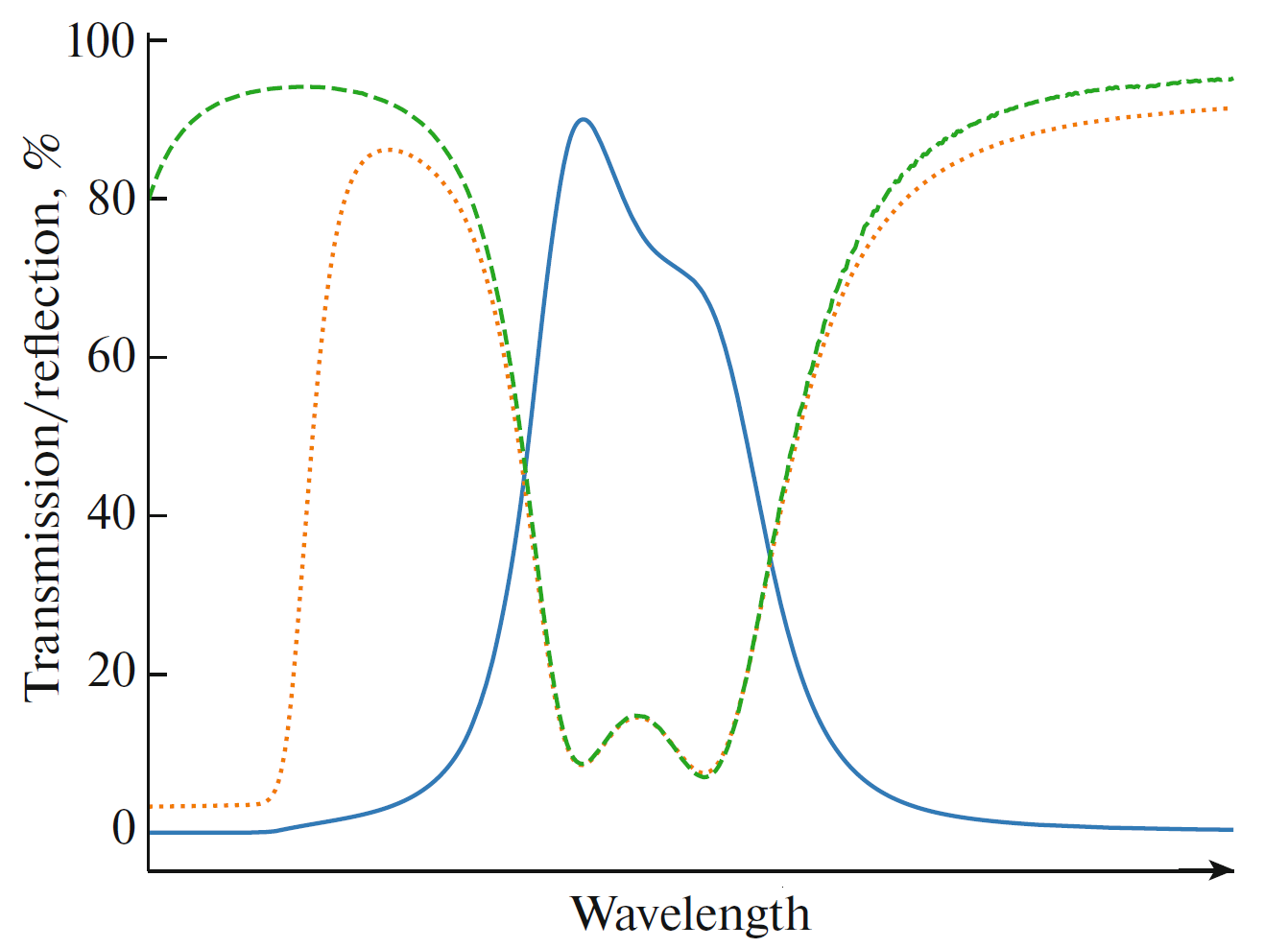}}
  \caption{The shape of the transmission (solid blue curve) and reflection curves from both sides (dashed line~--- looking at the filter from the detector side, dotted line~--- from the telescope side) for a typical interferometric filter}
  \label{fig:filter_reflection}
\end{figure}

In this formula, $\Omega_T$ is the total solid angle of the beam coming from the telescope, and $\varepsilon$ is the emissivity of the filter. If the filter were completely transparent (or completely mirrored), its emissivity $\varepsilon$ would be 0. A real filter will have a transmission curve $T(\lambda)$ determined by the desired photometric system (see Fig.~\ref{fig:atm_smooth}). In addition, the filters will have a reflectance curve $R(\lambda)$ when viewed from the detector side. Typical transmission and reflection curves of interference filters, are shown in Fig.~\ref{fig:filter_reflection}. Thus, the magnitude of the emissivity can be estimated by the formula $\varepsilon(\lambda) = 1-T(\lambda)-R(\lambda)$. It usually ranges from a few percent to 10-15 \% in the range of our interest.

Then, it is necessary to take into account the temperature of the radiation reflected from the filter. Due to the features of the device design described further in section \ref{sec_camera} in this calculation, we took it equal to the internal temperature of the detector cryostat $-100^\circ$C. This reflected flux turns out to be much smaller than the intrinsic radiation of the filters. The obtained values of fluxes from filters strongly depend on the chosen value of emissivity of the filter material. The results of calculations for the value $\varepsilon=0.04$ can be seen in Table ~\ref{table:filters}. The values for both filters are the same because it is considered that $\varepsilon$ does not depend on the spectral range.

\begin{table}[htbp]
\caption{Background produced by photometric filters (for $\varepsilon = 4\% $ over the entire sensitivity range of the detector)}
\begin{center}
\begin{tabular}{| l | l | }
\hline
Filter temperature, $^\circ$C & Background, ph/(s$\cdot$pix) \\
\hline
-20 & 730 000\\
-10 & 1 200 000\\
0 & 1 800 000  \\
10 & 2 600 000 \\
20 & 3 800 000 \\
\hline
\end{tabular}
\label{table:filters}
\end{center}
\end{table}

\section{Possible design of the IR camera}\label{sec_camera}
\mbox{}\vspace{-\baselineskip}

One of the main tasks to be solved when creating an infrared device is to minimize the instrumental background. This requires minimizing the number of warm optical elements in the instrument circuitry. In our case, the detector is shipped with its own cryostat and cooling system based on a Stirling machine. The low power of the cooler leads to the fact that only the detector has direct contact with the site cooled by the machine. The field of view of the detector is limited by a cold aperture located in one vacuumized volume with the detector. The entrance window is a thin germanium plate mounted on the end of a polished steel cone, which can be used to mount the camera. In this case, the camera can be mounted on an additional cryostat with a vacuumized volume in which to arrange the cooled filters, optics, and limiting cold apertures. However, the use of an additional cryostat dramatically increases the final cost of the project. Therefore, we propose a simplified camera design for consideration.

\begin{figure}[tbh!]
  \centerline{
  \includegraphics[width=0.6\linewidth]{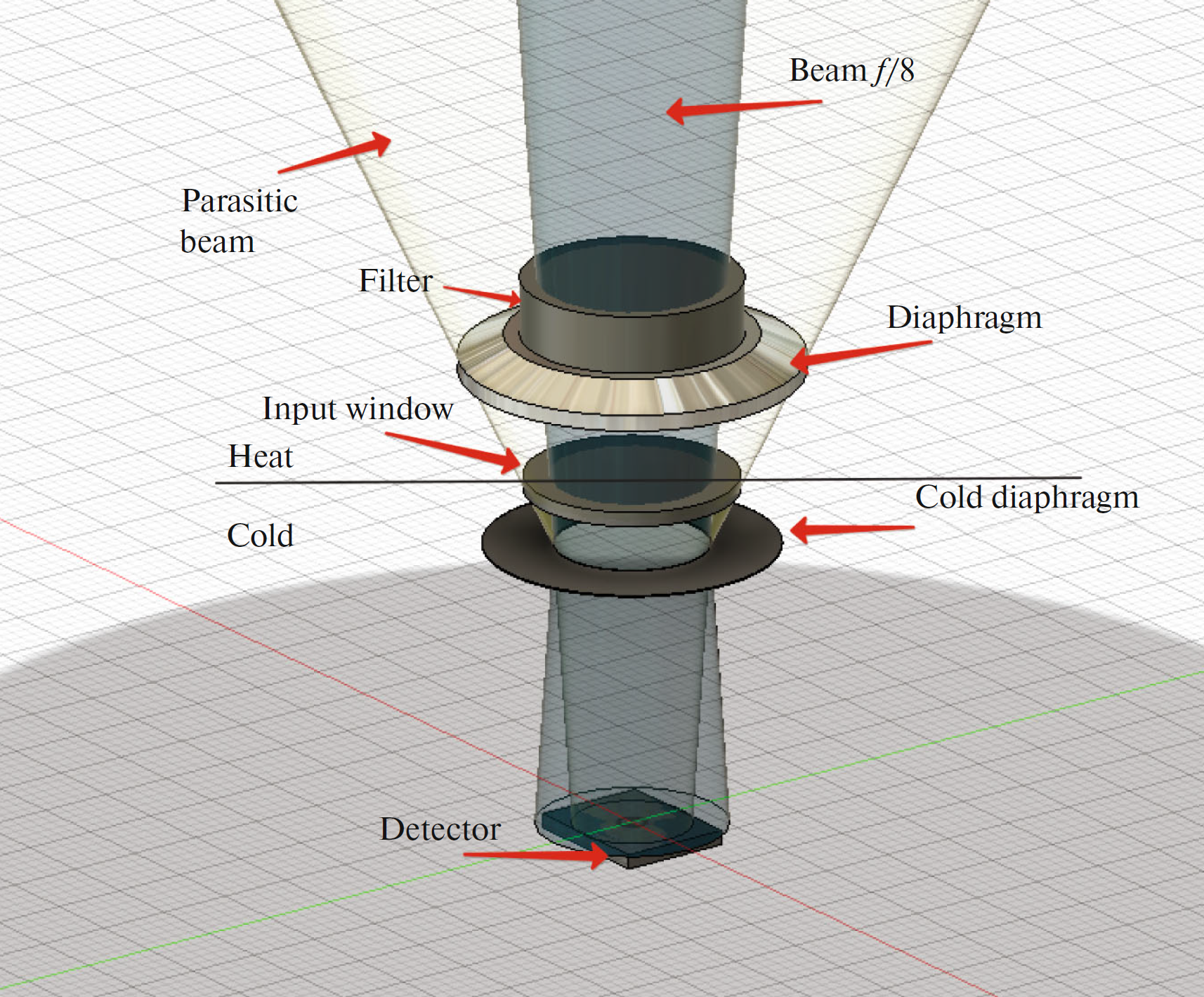}}
  \caption{Beam path, aperture and filter position in camera design}
  \label{fig:cam_filt}
\end{figure}

In a design without a cryostat, it is possible to limit to one element located in the light beam feeding the instrument, namely the photometric filter. The relative aperture of the 2.5-m telescope is $A=1/8$, which gives small ray convergence angles and allows the interference filters to operate in a converging beam without a noticeable change in transmission bandwidth. However, this angle is much smaller than the angular field of view of the detector, which is often calculated based on $A=1/4$. Outside of their transmission bandwidth, interference filters are mirrors with high reflection coefficient. Characteristic transmission and reflection curves can be seen in Fig.~\ref{fig:filter_reflection}. Thus, to reduce the instrumental background, it is necessary to minimize the temperature of the reflected radiation. By installing the filters in close proximity to the entrance window of the detector, it is possible to achieve that the light filters will reflect the cold interior of the detector. To limit the magnitude of the signal through the filter due to the difference in relative apertures of the telescope and detector, a mirrored aperture with an appropriate inner size and an outer size larger than the filter should be installed. The mirrored diaphragm should have a concave surface to exclude reflection of warm parts of the camera in it. The general scheme of the resulting assembly is shown on the figure \ref{fig:cam_filt}.

\section{Estimation of the limiting magnitude}
\mbox{}\vspace{-\baselineskip}

On the basis of the obtained data on the instrumental background and sky background in the analyzed ranges for the selected variant of the camera design, it is possible to estimate the limiting stellar magnitudes. For this purpose, let us write down the theoretical signal-to-noise ratio ($SNR$):

\begin{equation}
SNR =\frac{F_0\cdot2.512^{-m} \cdot t }{\sqrt{F_0\cdot2.512^{-m} \cdot t + 2((a_s/a_{pix})^2\cdot F_{bg} \cdot t + (a_s/a_{pix})^2 \cdot \sigma_d^2/n)}},
\label{eq:snr}
\end{equation}

where $t$ is the signal accumulation time, $F_0$ is the number of photoelectrons captured by the detector from a star with brightness $0^m$, $m$ is the stellar magnitude of the object, $F_{bg}$ is the number of photoelectrons detected by the detector from the background (the sum of the instrumental background, sky background, and dark current) per 1 square arcsec of the sky, $a_s$ and $a_{pix}$ are the dimensions of the star and pixel image in arcsec, respectively, $n$ is the number of single frames involved in obtaining the final image. The factor 2 before the second summand in the denominator accounts for the fact that subtracting the background during processing will double the corresponding noise. The calculations assume that the quantum efficiency of the detector is 0.85, and the transmission of the atmosphere, telescope, and filter is 0.5. Unfortunately, the manufacturer does not specify the values of the readout noise and dark current of the detector, but for such detectors the noise caused by these effects is usually much lower than the Poisson background noise\cite{2021RAA....21...81Z}. Under these conditions, the value of the expected limiting stellar magnitude for a given background level, $SNR$, and accumulation time can be easily obtained by solving the equation \ref{eq:snr} with respect to $m$.

Limitations on the minimum exposure time by the electronics of the detector affect the ability to observe bright objects. A typical value of the minimum exposure of commercial detectors can be considered to be 300-400~$\mu$s. With a cell capacity of $8\cdot10^6$~e- we can observe stars no brighter than $-4.5^m$ in $L'$ and $-7^m$ in $M'$.

The maximum duration of a single exposure is limited by the maximum pixel capacity ($8\cdot10^6$~e-). For the characteristic background values found above, the maximum exposure will be $t_{max}=1..3$~s. However, if the readout noise is negligible, one long exposure is not fundamentally distinguishable from the sum of many short exposures. This allows to bypass limitations on the value of $t_{max}$ and, if necessary, to accumulate hundreds of frames to obtain the necessary $SNR$. In addition, observations in the mode of stacking of separate frames to obtain the final image allow to organize the modulation of the light flux, when the center of the field of view moves by $5~- 10$~arcsec between separate frames. The modulation frequency is quite high ($0.3..10$~Hz) and the implementation of this method requires the use of an additional chopping mirror installed between the entrance window of the camera and the elements of the telescope optical scheme. For further calculations, we set the accumulation time $t$ equal to 1~s.

We will consider the value $SNR=3$ as the detection limit of a point object on the frame. In Table ~\ref{table:limits} we present the results of calculations for different cases. The first three rows of the table are obtained with fixed values of $\varepsilon=0.04$ and star image size $FWHM=1''$. It can be seen that the transition from a cold telescope and cold filters to heated filters leads to a loss of about $1^m$. Observations made at low image quality lead to the same loss. The use of filters with high emissivity also degrades the camera limit.

\begin{figure}[tbh!]
  \centerline{
  \includegraphics[width=0.6\linewidth]{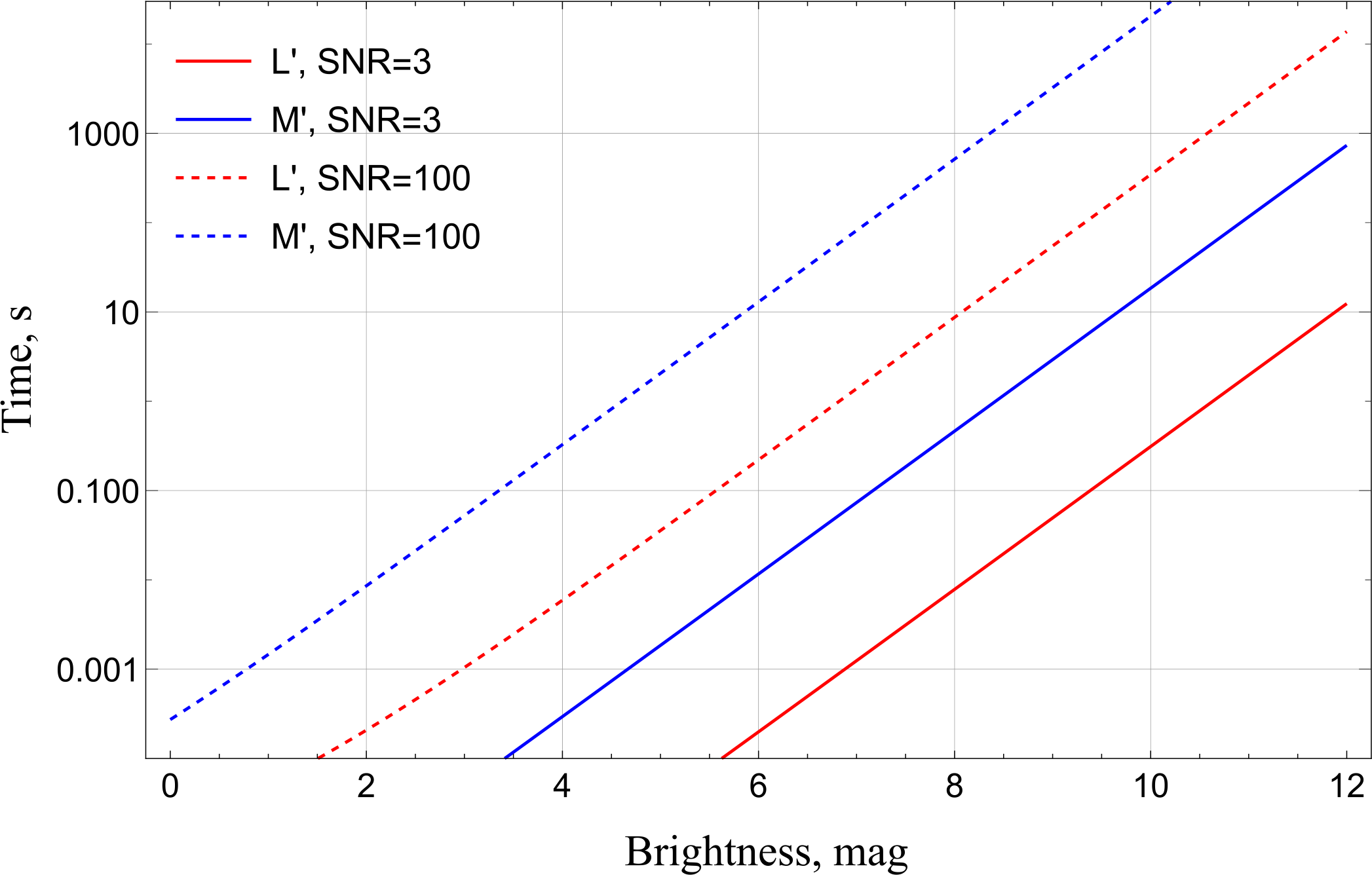}}
\caption{The dependence of exposure time on the brightness of the star at telescope temperature and filters $0^\circ$C, $FWHM = 1''$ and $\varepsilon_{fltr}=0.04$. The red solid line~--- $L'$ band for $SNR=3$. Red dashed line~--- $L'$ band for $SNR=100$. Blue solid line~--- $M'$ band for $SNR=3$. Blue dashed line~--- $M'$ band for $SNR=100$}
  \label{fig:snr_lm}
\end{figure}

In Fig.~\ref{fig:snr_lm}, the dependence of the integral accumulation time $t$ required to reach a certain stellar magnitude $m$ of a point object for two typical values of $SNR$ is given for the observing conditions specified in the second row of Table ~\ref{table:limits} for two typical values of $SNR$: $SNR=3$ is the limit of detectability and $SNR=100$ is the case of accurate photometry.

\begin{table}[htbp]
\caption{The estimated limiting stellar magnitudes of point sources}
\begin{center}
\begin{tabular}{| l | l | l | }
\hline
Conditions & $L'$ band & $M'$ band  \\
\hline
$T_{tel}=-20^\circ$C, $T_{fltr}=-40^\circ$C, $\varepsilon_{fltr}=0.04$, $FWHM=1''$ & 11.2$^m$ & 8.8$^m$\\
$T_{tel}=0^\circ$C, $T_{fltr}=0^\circ$C, $\varepsilon_{fltr}=0.04$, $FWHM=1''$ & 10.6$^m$ & 8.4$^m$\\
$T_{tel}=+20^\circ$C, $T_{fltr}=+20^\circ$C, $\varepsilon_{fltr}=0.04$, $FWHM=1''$ & 10.2$^m$ & 8.1$^m$\\
$T_{tel}=+20^\circ$C, $T_{fltr}=+20^\circ$C, $\varepsilon_{fltr}=0.2$, $FWHM=1''$ & 9.7$^m$ & 7.6$^m$\\
$T_{tel}=+20^\circ$C, $T_{fltr}=+20^\circ$C, $\varepsilon_{fltr}=0.04$, $FWHM=2.5''$ & 9.2$^m$ & 7.1$^m$\\
\hline
\end{tabular}
\label{table:limits}
\end{center}
\end{table}

\section*{Conclusion}
\mbox{}\vspace{-\baselineskip}

For the combination of a universal optical telescope and a commercial IR camera operating in the short-wave end of the mid-IR range (3-5~$\mu$m) considered in our work, for various observing conditions we obtained the values of the contribution to the background signal from all components: the Earth's atmosphere, the telescope, and the structural elements of the camera (see Tables ~\ref{table:telescope}, \ref{table:filters}, and Fig.~\ref{fig:sky_bg}). A comparison of the values of these components suggests that their contributions are roughly equal in the $L'$ band, while the $M'$ band is dominated by atmospheric radiation at telescope temperatures below $10^\circ$~C, and dominated by telescope radiation at warmer telescopes. To minimize the contribution from the camera components (while keeping the system cost to a minimum), we propose a layout of a filter with a concave mirrored aperture (Fig.~\ref{fig:cam_filt}).

Calculations of the limiting stellar magnitude of objects accessible to observations with the camera were carried out for different observing conditions. Table ~\ref{table:limits} shows that the losses during the transition from variant to variant are $0.5-1^m$. The greatest influence is the quality of the image during observations. In the paper\cite{2021RAA....21...81Z}, an estimate of the limiting magnitude in the $M'$ band ($M'=13^m$) was obtained for an adapted 10~m-diameter telescope installed in a high-altitude observatory for the same $SNR$ values and accumulation time. A direct conversion to the 2.5-m CMO telescope results in a limiting magnitude value of $10^m$, which is about $1^m$ better than the estimate we obtained for the proposed camera design. This difference is mainly due to the smaller contribution of the atmosphere at high altitude and the lower temperature of the telescope.

Our estimates of the limiting magnitude show that more than 1 million objects will be available for observations with the proposed infrared camera, according to the WISE\cite{WISE} catalog. These are stars of various types and classes, galaxies of various types, planetary nebulae, and other objects. It is easy to get estimates of the sizes of the Solar System bodies that can be accessible to observations with the described equipment. For example, an ''gray'' asteroid with a diameter of 100~km in the main belt in opposition with an albedo of 0.2 would have a brightness of $m_{L'} = 8.7^m$ and $m_{M'} = 8.5^m$, which is close to our detection limit. A body with a diameter of 30~m at the Moon's distance with the same albedo would have brightness $m_{L'} = 9.6^m$ and $m_{M'} = 7.7^m$.  This large change in the color index is due to the strongly increased intrinsic emission of the body closer to the Sun in the longer wavelength. If we consider low Earth orbit, the camera will have access to centimeter-sized objects.

\bigskip 

The study was supported by the Scientific Educational School of Lomonosov Moscow State University ``Fundamental and Applied Space Research''. The work of S. Zheltoukhov was supported by the Foundation for the Advancement of Theoretical Physics and Mathematics ``BASIS''. This publication makes use of data products from the Wide-field Infrared Survey Explorer, which is a joint project of the University of California, Los Angeles, and the Jet Propulsion Laboratory/California Institute of Technology, funded by the National Aeronautics and Space Administration.

\bigskip\footnotesize

\end {document}